\documentclass{article}
\usepackage{epsfig}

\def\be{\begin{equation}}
\def\ee{\end{equation}}
\def\bea{\begin{eqnarray}}
\def\eea{\end{eqnarray}}
\def\ket#1{\hbox{$\vert #1\rangle$}}   
\def\bra#1{\hbox{$\langle #1\vert$}}   
\def\oneh{{\textstyle {1\over 2}}}
\def\onet{{\textstyle {1\over 3}}}

\def          
\simeq{{\ \lower2pt\hbox{$-$}\mkern-13mu \raise2pt \hbox{$\sim$}\ }} 
\def\dirac#1{\slash \mkern-10mu #1}

\begin{document}

\begin{center}
\Large{\bf Linking generalized parton distributions to constituent quark
models}
\end{center}

\bigskip

\begin{center}
\large S.~Boffi$^a$, B.~Pasquini$^b$,$^c$, M.~Traini$^c$
\end{center}

\medskip

\begin{center}
$^a$ {Dipartimento di Fisica Nucleare e Teorica, Universit\`a degli
Studi di Pavia and INFN, Sezione di Pavia, Pavia, Italy}
\end{center}
\begin{center}
$^b$ {ECT$^*$, Villazzano (Trento), Italy}
\end{center}
\begin{center}
$^c$ {Dipartimento di Fisica, Universit\`a degli Studi di Trento, Povo
(Trento), and INFN, Sezione di Trento, Trento, Italy}
\end{center}

\bigskip

\begin{abstract}
The link between the nucleon generalized parton distributions and the
non-diagonal one-body density matrix in momentum space is studied. Attention is
focussed on the region where quark generalized parton distributions (GPD's)
describe emission and reabsorption of a single active quark by the target
nucleon. The correct covariant connection with wave functions used in any
constituent quark model is established. Results obtained with different
constituent quark models are presented for the unpolarized quark GPD's. 
\end{abstract}

\hspace{0.5cm}\small Key words:
generalized parton distributions, constituent quark models

\hspace{0.5cm}PACS 12.39.-x, 13.60.Fz, 13.60.Hb, 14.20.Dh

\normalsize


\section{Introduction}

In recent years much theoretical activity has been devoted to generalized parton
distributions (GPD's). They are defined as nondiagonal hadronic
matrix elements of bilocal products of the light-front quark and gluon field
operators and they interpolate between the inclusive physics of parton
distributions and the exclusive limit of electromagnetic form
factors~\cite{muller,radyushkin96a,radyushkin96b,ji78,radyushkin97,diehl01}. As 
such they contain information on parton correlations and the internal spin
structure of the nucleon. But the main interest to study GPD's comes from the
fact that, due to the factorization
theorem~\cite{radyushkin97,jiosborne,collins99}, they are candidate to provide
us with a unifying theoretical background suitable to describe a variety of
inclusive and exclusive processes in the deep inelastic scattering (DIS) regime.
In particular, they enter the cross section for exclusive photon production,
i.e. deeply virtual Compton scattering
(DVCS)~\cite{ji78,radyushkin97,ji55,belitsky}, and hard diffractive
electroproduction of (longitudinal) vector and pseudo-scalar
meson~\cite{collins97}. In addition, in the forward limit they become diagonal
matrix elements giving the usual DIS parton distribution, and their first moment
gives the nucleon elastic form factors~\cite{ji78}. Basic properties of the
GPD's have been reviewed in Refs.~\cite{jig,radyushkin01,goeke}.  

Much effort is underway related to the measurement of these functions. A general
discussion about the problems arising in the experimental study of the
GPD's and the first experimental evidences have been presented in
Ref.~\cite{guidal}. Prospects for future measurements of DVCS are studied in
Ref.~\cite{nowak}. On the other hand, model calculations are a necessary 
prerequisite to define suitable strategies to extract GPD's from experiments. 
In the literature there are two approaches used to model the nucleon GPD's. 
One is a phenomenological construction based on reduction formulas where GPD's
are related to the usual parton distributions by factorizing the momentum
transfer dependence due to the nucleon electroweak form
factors~\cite{radyushkin99,radyushkin01,marc}.
This leads to double distribution functions parametrizing the nonforward matrix
elements involved in DVCS and hard exclusive electroproduction processes
discussed, e.g., in Ref.~\cite{vanderhaeghen}. Another approach is based on 
direct calculation of GPD's in specific dynamical models. The first model
calculations were performed using the MIT bag model~\cite{bag}. Further
calculations have been performed in the chiral quark-soliton
model~\cite{petrov,penttinen}. 

A complete and exact overlap representation of GPD's has been recently worked
out within the framework of light-cone 
quantisation~\cite{diehl,brodsky}. In particular, quark GPD's are obtained as
overlaps of light-cone wave functions (LCWF's) indicating that they are deeply
connected with the non-diagonal one-body density matrix in momentum space
occurring in any nonrelativistic many-body problem. A preliminar investigation
of such an approach was presented in Ref.~\cite{diehl99}.

In fact, direct calculation of LCWF's from first principles is a difficult task.
On the other hand, constituent quark models (CQM's) have been quite successful in
describing the spectrum of hadrons and their low-energy dynamics. Therefore it
is interesting to explore the connection between GPD's and CQM wave functions at
least in the allowed kinematic range where only quark degrees of freedom are
effective. This corresponds to the kinematic region where GPD's describe how a
quark is taken out from the proton and, having undergone a
hard scattering, is inserted back as a quark inside the scattered proton.
This connection has been studied in Ref.~\cite{scopetta} where a simple
nonrelativistic quark model has been used to calculate one of the twist-two
GPD's at a low-energy scale. However, as in the case of parton distributions
calculated from CQM wave functions a nonrelativistic approach is obviously
insufficient~\cite{faccioli} and a consequence of it is that the obtained GPD's
are not defined on their natural support and the particle number and momentum
sum rules are not always preserved. The support violation occurs also with the
MIT bag 
model~\cite{bag} because the initial and final nucleons are not good momentum
eigenstates. These are common problems arising also in the case
of parton distributions calculated from CQM wave functions~\cite{conci,thomas}
where they were cured by appropriately taking into account translation
invariance~\cite{mair}. 

CQM's rely on quantum theory with a finite number of degrees of freedom. In this
case relativity can be incorporated quite naturally by utilizing relativistic
Hamiltonian dynamics~\cite{keister} and the Bakamjian-Thomas~\cite{BT}
construction of the Hamiltonian for a system of interacting particles. In such an
approach CQM wave functions can be considered as eigenfunctions of the
nucleon Hamiltonian in the instant-form dynamics and can simply be related to
wave functions in any form of relativistic Hamiltonian dynamics. Examples are the
light-front description of electromagnetic form factors (see Ref.~\cite{pace}
and references therein) and
the covariant calculation of the nucleon electroweak form factors~\cite{robert}
in a chiral CQM~\cite{graz} with the point-form approach. Similarly, one can
obtain a link between CQM wave functions and LCWF's with the corresponding
transformation from the instant-form to the front-form representation as it is
here proposed. GPD's in the allowed kinematic region are then obtained in a
covariant approach and they exhibit the exact forward limit reproducing the
parton distribution with the correct support and automatically fulfilling the
particle number and momentum sum rules.

The paper is organized as follows. In Sect. 2 the relevant definitions and
properties of GPD's in terms of LCWF's are recalled. In Sect. 3 the connection
is established between the (light-cone) front-form and the (canonical)
instant-form description. This connection is used  in Sect. 4 to obtain the
unpolarized GPD's and their forward limits, i.e. the usual DIS parton
distributions in terms of CQM wave functions. The results obtained with
different CQM wave functions are presented in Sect. 5, confining some technical
details in the Appendix, and concluding remarks are given in Sect. 6. 


\section{The unpolarized generalized parton distributions}

\begin{figure}[h]
\begin{center}
\epsfig{file=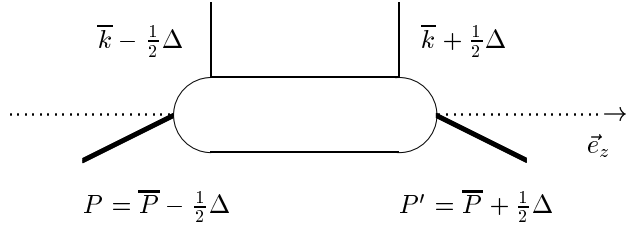}
\end{center}
\caption{\small The symmetric frame of reference.}
\label{fig:symmetric}
\end{figure}

For definiteness let us consider virtual Compton scattering where a lepton
exchanges a virtual photon of momentum $q^\mu$ with a nucleon of momentum
$P^\mu$, producing a real photon of momentum ${q'}^\mu$ and a recoil nucleon of
momentum ${P'}^\mu$. Ultimately we are here interested in kinematic conditions
similar to those familiar in the DIS regime that is
characterized by the Bjorken limit, i.e. $Q^2=-q^2\to\infty$, $P\cdot
q\to\infty$, and fixed $x_B=Q^2/2(P\cdot q)$. When focusing on the deeply
virtual kinematic regime of $q^\mu$, i.e. 
DVCS, a generalization of the Bjorken kinematics is considered, namely the
c.m. energy $s=(P+q)^2$ and the photon virtuality $Q^2$ are large while the
invariant momentum square $t=\Delta^2=2P\cdot\Delta$ is
small~\cite{muller,ji78,radyushkin97}. Under these conditions the factorization 
theorem~\cite{radyushkin97,jiosborne,collins99} tells us that the amplitude
factorizes in a hard scattering part (which is exactly calculable in 
perturbative QCD) and a soft, nonperturbative nucleon structure part. The
contribution of the hard scattering part to leading order corresponds to the
so called handbag diagram with photon scattering on a single parton (quark or
antiquark). Consequently, the soft part is a quark-quark correlation function,
representing the process where a parton is taken out of the initial nucleon and
reinserted back into the final nucleon after hard scattering. To describe such a
process, according to Ref.~\cite{ji78} it is useful to choose a symmetric frame
of reference where the virtual photon momentum $q^\mu$ and the average nucleon
momentum $\overline P^\mu=\oneh(P^\mu+{P'}^\mu)$ are collinear along the $z$
axis and in opposite  directions (Fig.~\ref{fig:symmetric}). It is also useful
to use the component notation $a^\mu = [a^+,a^-,\vec{a}_\perp]$ for any
four-vector $a^\mu$ with light-cone components $a^\pm=(a^0\pm a^3)/\sqrt{2}$ and the
transverse part $\vec{a}_\perp =(a^1,a^2)$.  

In the following we only consider the case of unpolarized quarks inside the
nucleon. Then for each flavor $q$ the soft amplitude in Fig.~\ref{fig:symmetric}
becomes  
\be
\label{eq:definition}
F^q_{\lambda'\lambda}(\overline x,\xi,t) =
\left. \frac{1}{4\pi}\int dy^-\, e^{i\overline x\overline P^+y^-}
\bra{P',\lambda'}\overline\psi(-\oneh y)\,\dirac n\, \psi(\oneh y)\ket{P,\lambda}
\right\vert_{y^+=\vec{y}_\perp=0},
\ee
where $n$ is a lightlike vector proportional to $(1,0,0,-1)$, 
$\lambda$ ($\lambda'$) is the helicity of the initial (final) nucleon and
the quark-quark correlation function is integrated along the light-cone
distance $y^-$ at equal light-cone time ($y^+=0$) and at zero transverse
separation ($\vec{y}_\perp=0$) between the quarks. The resulting
one-dimensional Fourier integral along the light-cone distance $y^-$ is with
respect to the quark light-cone momentum $\overline k^+=\overline x\overline
P^+$.
The so called skewedness parameter $\xi$ describes the longitudinal change
of the nucleon momentum, $2\xi = -\Delta^+/\overline P^+$. The link operator
normally needed to make the definition~(\ref{eq:definition}) gauge invariant
does not appear because we also choose the gauge $A^+=0$ which reduces the link
operator to unity.

Following Ref.~\cite{ji78} the leading twist (twist-two) part of this amplitude
can be parametrized as 
\bea
F^q_{\lambda'\lambda}(\overline x,\xi,t) 
& = & \frac{1}{2\overline P^+}
\, \overline u(P',\lambda')\,\gamma^+\, u(P,\lambda) \, H^q(\overline x,\xi,t)
 \nonumber\\
& & \quad + \frac{1}{2\overline P^+} 
\, \overline u(P',\lambda')\,
\frac{i\sigma^{+\nu}\Delta_\nu}{2M}  \,u(P,\lambda)\,E^q(\overline x,\xi,t) ,
\\ \nonumber
\eea
where $u(P,\lambda)$ is the nucleon Dirac spinor and $H^q(\overline
x,\xi,t)$ and  $E^q(\overline x,\xi,t)$ are the chiral-even (helicity conserving)
and chiral-odd (helicity flip) GPD's for partons of flavor $q$, respectively.

An explicit expression in term of LCWF's can be obtained following the lines of
Refs.~\cite{diehl,brodsky}. The hadronic state is written as 
\be
\label{eq:hadronstate}
\ket{P,\lambda}=\sum_{N,\beta}
\int\left[\frac{d\overline x}{\sqrt{\overline x}}\right]_N [d\vec{k}_\perp]_N
\Psi_{\lambda,N,\beta}(r)\ket{N,\beta; k_1,\dots,k_N},
\ee
where the integration measures are defined by
\be
\label{eq:measone}
\left[\frac{d\overline x}{\sqrt{\overline x}}\right]_N 
= \left(\prod_{i=1}^N \frac{d\overline x_i}{\sqrt{\overline x_i}}\right)
\delta\left(1-\sum_{i=1}^N \overline x_i\right),
\ee
\be
\label{eq:meastwo}
[d\vec{k}_\perp]_N = \left(\prod_{i=1}^N
\frac{d\vec{k}_{\perp\,i}}{2(2\pi)^3}\right)\,2(2\pi)^3\,
\delta\left(\sum_{i=1}^N \vec{k}_{\perp\,i} - \vec{P}_\perp\right).
\ee
The function $\Psi_{\lambda,N,\beta}(r)$ is the momentum LCWF of the $N$-parton
Fock state $\ket{N,\beta; k_1,\dots,k_N}$ (also including gluons), with $\beta$
indicating flavour, helicity and colour quantum numbers. The LCWF's depend on
the momentum coordinates (collectively indicated by $r$) of the partons relative
to the hadron momentum. The $N$-parton states are normalized as
\bea
& & \langle N',\beta'; k'_1,\dots,k'_N \ket{N,\beta; k_1,\dots,k_N} 
\nonumber \\
& & \qquad =
\delta_{N'N}\,\delta_{\beta'\beta}\prod_{i=1}^N 2k^+_i(2\pi)^3 
\delta({k'_i}^+-k_i^+)\,\delta({{\vec{k}'_{\perp\,i}}}-\vec{k}_{\perp\,i}).
\\ \nonumber
\eea
Correspondingly, the hadron states are covariantly normalized as
\be
\bra{P',\lambda'}P,\lambda\rangle = 2P^+(2\pi)^3\,
\delta({P'}^+-P^+)\,\delta(\vec{P}'_\perp -\vec{P}_\perp)
\,\delta_{\lambda'\lambda}, 
\ee
with
\be
\sum_{N,\beta}\int[d\overline x]_N[d\vec{k}_\perp]_N
\vert\Psi_{\lambda,N,\beta}(r)\vert^2 = 1.
\ee

Having in mind the link between GPD's and CQM wave functions we have to confine our discussion to the region $\xi<\overline x<1$. In this region
and in the symmetric frame
\bea
\label{eq:base}
F^q_{\lambda'\lambda}(\overline x,\xi,t) & = &
\sum_{N,\beta}\left(\sqrt{1-\xi}\right)^{2-N} \left(\sqrt{1+\xi}\right)^{2-N}
\sum_{j=1}^N \delta_{s_jq}
\nonumber \\
& & \quad\times
\int[d\overline x]_N[d\vec{k}_\perp]_N\,\delta(\overline x-\overline x_j)
\Psi^{*}_{\lambda',N,\beta}(r') \Psi_{\lambda,N,\beta}(r)\Theta(\overline x_j),
\\ \nonumber
\eea
where $s_j$ labels the quantum numbers of the $j$th parton, $\beta$ specifies
all other quantum necessary for the $N$-parton state, and the set of kinematic
variables $r,r'$ are defined as follows: for the final struck quark,
\be
\label{eq:finalconf}
y'_j = \frac{\overline k^+_j + \oneh\Delta^+}{\overline P^+ + \oneh\Delta^+}
= \frac{\displaystyle \overline x_j-\xi}{\displaystyle 1-\xi},\qquad
\vec{\kappa}'_{\perp j} = \vec{k}_{\perp j} + \oneh
\frac{\displaystyle 1-\overline x_j}{\displaystyle 1-\xi}\vec{\Delta}_\perp, 
\ee
for the final $N-1$ spectators ($i\ne j$),
\be
y'_i = \frac{\displaystyle \overline x_i}{\displaystyle 1-\xi},
\ \vec{\kappa}'_{\perp i}= \vec{k}_{\perp i} - \oneh
\frac{\displaystyle \overline x_i}{\displaystyle 1-\xi}\vec{\Delta}_\perp,
\ee
and for the initial struck quark
\be
\label{eq:initialconf}
y_j = \frac{\overline k^+_j - \oneh\Delta^+}{\overline P^+ - \oneh\Delta^+}
= \frac{\displaystyle \overline x_j+\xi}{\displaystyle 1+\xi}, 
\qquad
\vec{\kappa}_{\perp j}= \vec{k}_{\perp j} - \oneh
\frac{\displaystyle 1-\overline x_j}{\displaystyle 1+\xi}\vec{\Delta}_\perp,
\ee
for the initial $N-1$ spectators ($i\ne j$),
\be
y_i = \frac{\displaystyle \overline x_i}{\displaystyle 1+\xi},
\ \vec{\kappa}_{\perp i}=\vec{k}_{\perp i} + \oneh
\frac{\displaystyle \overline x_i}{\displaystyle 1+\xi}\vec{\Delta}_\perp.
\ee
Working out the spinor products we have
\bea
F^q_{++}(\overline x,\xi,t) & = & F^q_{--}(\overline x,\xi,t) \nonumber\\
\label{eq:noflip}
& = &\sqrt{1-\xi^2}\,H^q(\overline x,\xi,t) 
-\frac{\xi^2}{\sqrt{1-\xi^2}}\, E^q(\overline x,\xi,t),\\
\label{eq:flippa}
F^q_{-+}(\overline x,\xi,t) & = & -\left(F^q_{+-}(\overline x,\xi,t) \right)^*
= \eta\frac{\sqrt{t_0-t}}{2M} \, E^q(\overline x,\xi,t),
\\ \nonumber
\eea
where
\be
\eta=\frac{\Delta^1 + i\Delta^2}{\vert\vec{\Delta}_\perp\vert},
\ee
and
\be
\label{eq:minimalt}
- t_0 = \frac{4{\xi}^2M^2}{1-\xi^2}
\ee
is the minimal value for $-t$ at given $\xi$.

Therefore, one can extract $H^q$ and $E^q$ separately from the knowledge of
$F^q_{\lambda'\lambda}$. In particular, $E^q$ is directly given by
Eq.~(\ref{eq:flippa}), and
\be
H^q(\overline x,\xi,t) = \frac{1}{\sqrt{1-\xi^2}}
\left[ F^q_{++}(\overline x,\xi,t)
+\frac{2M\xi^2}{\eta\,\sqrt{t_0-t}\sqrt{1-\xi^2}}\, 
F^q_{-+}(\overline x,\xi,t)\right].
\ee


\section{Nucleon wave functions in front and instant form}

In this section a connection will be established between the (light-cone)
front-form and the (canonical) instant-form description.  

When necessary, labels $[f]$ and $[c]$ on the wave function will refer to
front form and canonical form, respectively, and vectors with tilde are defined
in the front form. 

\subsection{Single-parton states}

Omitting colour degrees of freedom that do not matter
in the following considerations, single-parton states can be defined either with
front-form coordinates, 
\be
\ket{\vec{\tilde{k}},\lambda,\tau}_{[f]}, \qquad \vec{\tilde{k}} =(k^+,\vec{k}_\perp),
\ee 
(with isospin $\tau$) or with instant-form coordinates, 
\be
\ket{\vec{k},\lambda,\tau}_{[c]}, \qquad \vec{k} = (k_x, k_y, k_z),
\ee
with corresponding normalizations
\bea
\label{eq:normalizza}
_{[f]}\langle \vec{\tilde{k}'},\lambda',\tau'
\ket{\vec{\tilde{k}},\lambda,\tau}_{[f]}
& = & 2k^+(2\pi)^3\delta(k^+-{k'}^+)\,\delta(\vec{k}_\perp -\vec{k}'_\perp)
\,\delta_{\lambda\lambda'}\,\delta_{\tau\tau'},
\nonumber\\
_{[c]}\langle \vec{k}',\lambda',\tau'\ket{\vec{k},\lambda,\tau}_{[c]}
& = & \delta(\vec{k} - \vec{k}')
\,\delta_{\lambda\lambda'}\,\delta_{\tau\tau'}.
\\ \nonumber
\eea

The transformation from the instant- to the front-form representation reads
\be
\label{eq:connect}
\ket{\vec{\tilde{k}},\lambda,\tau}_{[f]} = \sum_{\lambda'\tau'}\int d\vec{k}'
\ket{\vec{k}',\lambda',\tau'}_{[c]}\, _{[c]}\bra{\vec{k}',\lambda',\tau'}
{\vec{\tilde{k}},\lambda,\tau}\rangle_{[f]},
\ee
where, according to Eq.~(4.35) of Ref.~\cite{keister},
\be
\label{eq:cfnon}
_{[c]}\bra{\vec{k}',\lambda',\tau'}{\vec{\tilde{k}},\lambda,\tau}\rangle_{[f]}
=\sqrt{2\omega}(2\pi)^{3/2}\,\delta(\vec{k}-\vec{k}')
{D}^{1/2}_{\lambda'\lambda}(R_{cf}(\vec{\tilde k}))\,\delta_{\tau\tau'}.  
\ee
Here $R_{cf}$ is a Melosh rotation, and the normalization constant
$\sqrt{2\omega}(2\pi)^{3/2}$ with $\omega\equiv k_0 = (k^+ + k^-)/\sqrt{2}$
derives from a different normalization used in Ref.~\cite{keister}. 

Then Eq.~(\ref{eq:connect}) becomes
\be
\label{eq:transcf}
\ket{\vec{\tilde{k}},\lambda,\tau}_{[f]}
=\sqrt{2\omega}(2\pi)^{3/2}\sum_{\lambda'}
{D}^{1/2}_{\lambda'\lambda}(R_{cf}(\vec{\tilde k}))
\ket{\vec{k},\lambda',\tau}_{[c]}.
\ee

\subsection{N-parton states in the front form}

In the front-form description the $N$-parton contribution
$\ket{N;\vec{\tilde{P}},\lambda}_{[f]}$ to the nucleon wave function can be
derived from the zero-momentum state by a suitable boost $L_f(Q)$ with
$Q=\vert\vec{P}\vert/M_0$ ($M_0$ being the mass of the noninteracting $N$-parton
system) and  
\be
L_f(Q)^\mu_\nu (1,0,0,0)^\nu = \frac{P^\mu}{M_0}.
\ee
In fact, under a Lorentz transformation $\Lambda$ we have
\be
U(\Lambda) \ket{N;\vec{\tilde{P}},\lambda}_{[f]} = \sum_{\lambda'}
\ket{N;\Lambda\vec{\tilde{P}},\lambda'}_{[f]}
\bra{\lambda'}R_W(\Lambda,P)\ket{\lambda},
\ee
where $R_W(\Lambda,P)$ is the Wigner rotation associated with the $\Lambda$
transformation. Assuming for $\Lambda$ the boost $\Lambda=L_f^{-1}(Q)$,
the Wigner rotation becomes the identity, $R_W(L_f^{-1}(Q),P)=\vec{1}$. Thus
\be
U(L_f(Q))\ket{N;\vec{\tilde{0}},\lambda}_{[f]}
=\ket{N;\vec{\tilde{P}},\lambda}_{[f]},
\ee
\be
U(L^{-1}_f(Q))\ket{N;\vec{\tilde{P}},\lambda}_{[f]}
=\ket{N;\vec{\tilde{0}},\lambda}_{[f]} .
\ee
In turn, according to Eq.~(\ref{eq:transcf}),
\be
\label{eq:fcrel}
\ket{N;\vec{\tilde{0}},\lambda}_{[f]} =
\sqrt{2M_0}(2\pi)^{3/2}\ket{N;\vec 0,\lambda}_{[c]}.
\ee
On the other hand, according to Eq.~(\ref{eq:hadronstate}) we have
\be
\label{eq:zerostate}
\ket{N;\vec{\tilde{0}},\lambda}_{[f]} = \sum_{\tau_i,\lambda_i}
\int\left[\frac{d\overline x}{\sqrt{\overline x}}\right]_N
[d\vec{k}_\perp]_N
\Psi_\lambda^{[f]}(\{\overline x_i,\vec{k}_{\perp i};\lambda_i,\tau_i\})
\prod_{i=1}^N \ket{\vec{\tilde{k}}_i,\lambda_i,\tau_i}_{[f]}.
\ee

\subsection{Connection with the instant-form representation}

In order to establish the connection with the instant-form representation one
has to work out the delta functions and the measures (Eqs.~(\ref{eq:measone})
and (\ref{eq:meastwo})). Since $\sum_i\omega_i
=M_0$, we have
\be
\delta\left(1-\sum_{i=1}^N \overline x_i\right)=
\delta\left(1- \sum_{i=1}^N\frac{k_{zi}+\omega_i}{M_0}\right) 
= M_0\,\delta\left(\sum_{i=1}^N k_{zi}\right),
\ee
\be
\label{eq:deltacf}
\delta\left(1-\sum_{i=1}^N \overline x_i\right)\,
\delta\left(\sum_{i=1}^N \vec{k}_{\perp i} \right)
= M_0\, \delta\left(\sum_{i=1}^N \vec{k}_i\right).
\ee
In addition,
\be
\prod_{i=1}^N  d\overline x_i \,d\vec{k}_{\perp\,i} = \prod_{i=1}^N d\vec{k}_i
\left\vert\frac{\partial(\overline x_i,\vec{k}_{\perp i})}{\partial(\vec{k}_i)}
\right\vert 
=\prod_{i=1}^N d\vec{k}_i \frac{\partial \overline x_i}{\partial k_{zi}} =
\prod_{i=1}^N d\vec{k}_i  \frac{\overline x_i}{\omega_i}.
\ee

We can then rewrite the zero-momentum state (\ref{eq:zerostate}) as
\bea
\label{eq:zeroff}
\ket{N;\vec{\tilde{0}},\lambda}_{[f]} &= & 
\sum_{\tau_i,\lambda_i} \int
\left[\prod_{i=1}^N d\vec{k}_i \frac{\overline x_i}{\omega_i} \frac{1}{\sqrt{\overline x_i}}\,
\frac{1}{2(2\pi)^3}\right] 2(2\pi)^3 M_0\, 
\delta\left(\sum_{i=1}^N \vec{k}_i\right)
\nonumber\\
& & \qquad \times
\Psi_\lambda^{[f]}(\{\overline x_i,\vec{k}_{\perp i};\lambda_i,\tau_i\})\prod_{i=1}^N
\sqrt{2\omega_i}(2\pi)^{3/2}\nonumber\\
& & \qquad\times\sum_{\mu_i}{D}^{1/2}_{\mu_i\lambda_i}(R_{cf}({\vec{\tilde k}}_i))
\ket{\vec{k}_i,\mu_i,\tau_i}_{[c]}
\nonumber\\
& =&  \sum_{\tau_i,\lambda_i} \int \left[\prod_{i=1}^N d\vec{k}_i  
\left(\frac{ \overline x_i}{\omega_i}\right)^{1/2}\right]
\,M_0\,\delta\left(\sum_{i=1}^N \vec{k}_i\right)
\Psi_\lambda^{[f]}(\{\overline x_i,\vec{k}_{\perp i};\lambda_i,\tau_i\})
\nonumber\\
& & \quad\times 2(2\pi)^3\prod_{i=1}^N
[2(2\pi)^3]^{-1/2}\sum_{\mu_i}{D}^{1/2}_{\mu_i\lambda_i}(R_{cf}({\vec{\tilde
k}}_i))
\ket{\vec{k}_i,\mu_i,\tau_i}_{[c]}.
\\ \nonumber
\eea
Therefore, the zero-momentum state in the canonical representation,
\be
\label{eq:zerocanon}
\ket{N;\vec 0,\lambda}_{[c]}
 = \sum_{\tau_i,\lambda_i} \int \prod_{i=1}^N d\vec{k}_i \, 
\delta\left(\sum_{i=1}^N \vec{k}_i\right)
\Psi_\lambda^{[c]}(\{\vec{k}_i;\lambda_i,\tau_i\})
\prod_{i=1}^N\ket{\vec{k}_i,\lambda_i,\tau_i}_{[c]},
\ee
can be rewritten in the light-front representation using the definition
(\ref{eq:fcrel}) and the result (\ref{eq:zeroff}) as
\bea
\label{eq:zerofront}
\ket{N;\vec 0,\lambda}_{[c]}
& = & \sum_{\tau_i,\lambda_i} \int \left[\prod_{i=1}^N d\vec{k}_i \, 
\left(\frac{\overline x_i}{\omega_i}\right)^{1/2} \right]\,\sqrt{M_0 }\,
\delta\left(\sum_{i=1}^N \vec{k}_i\right)
\Psi_\lambda^{[f]}(\{\overline x_i,\vec{k}_{\perp i};\lambda_i,\tau_i\})
\nonumber\\
& & \, \times[2(2\pi)^3]^{1/2}\prod_{i=1}^N
[2(2\pi)^3]^{-1/2}\sum_{\mu_i}{D}^{1/2}_{\mu_i\lambda_i}(R_{cf}({\vec{\tilde
k}}_i))
\ket{\vec{k}_i,\mu_i,\tau_i}_{[c]}.
\\ \nonumber
\eea

From Eqs.~(\ref{eq:zerocanon}) and (\ref{eq:zerofront}) a relationship can be
obtained between hadron wave functions $\Psi_\lambda^{[c]}$ and
$\Psi_\lambda^{[f]}$. This relation is here worked out in the case of a nucleon
with only three valence quarks, i.e. $N=3$. The three-quark (valence) Fock state
of the nucleon has been shown in Ref.~\cite{dziemb} to have only one independent
LCWF for all configurations where the quark helicities add up to the helicity of
the nucleon. This state is also known to almost exhaust the contribution to
GPD's at large values of $\overline x$~\cite{diehl99}.

By definition we have
\bea
& & \Psi_\lambda^{[c]}(\vec{k}_1,\vec{k}_2,\vec{k}_3;
\mu_1,\tau_1,\mu_2,\tau_2,\mu_3,\tau_3) \nonumber\\
& & \qquad = 
\bra{\vec{k}_1,\mu_1,\tau_1;\vec{k}_2,\mu_2,\tau_2;\vec{k}_3,\mu_3,\tau_3}
N;\vec{0},\lambda\rangle_{[c]}.
\eea
The bracket on the r.h.s. can be expressed in terms of the light-front
representation (\ref{eq:zerofront}), so that
\bea
\label{eq:psicf}
& &
\Psi_\lambda^{[c]}
(\vec{k}_1,\vec{k}_2,\vec{k}_3;\mu_1,\tau_1,\mu_2,\tau_2,\mu_3,\tau_3)
= \frac{1}{2(2\pi)^3}
\left[M_0\frac{\overline x_1\overline x_2\overline x_3}
{\omega_1\omega_2\omega_3}\right]^{1/2}
\nonumber \\ 
& & \qquad\times
\sum_{\lambda_1\lambda_2\lambda_3}
{D}^{1/2}_{\mu_1\lambda_1}(R_{cf}(\vec{\tilde k}_1))
{D}^{1/2}_{\mu_2\lambda_2}(R_{cf}(\vec{\tilde k}_2))
{D}^{1/2}_{\mu_3\lambda_3}(R_{cf}(\vec{\tilde k}_3))
\nonumber\\
& & \qquad\times
\Psi_\lambda^{[f]}(\overline x_1,\vec{k}_{\perp 1},\overline x_2,\vec{k}_{\perp
2},\overline x_3,\vec{k}_{\perp 3};\lambda_1,\tau_1,\lambda_2,\tau_2,\lambda_3,\tau_3). 
\\ \nonumber
\eea
Viceversa,
\bea
\label{eq:psifc}
& & \Psi_\lambda^{[f]}
(\overline x_1,\vec{k}_{\perp 1},
 \overline x_2,\vec{k}_{\perp 2},
 \overline x_3,\vec{k}_{\perp 3};
\lambda_1,\tau_1,\lambda_2,\tau_2,\lambda_3,\tau_3)  
\nonumber \\ 
& & \quad = 2(2\pi)^3
\left[\frac{1}{M_0}\frac{\omega_1\omega_2\omega_3}
{\overline x_1\overline x_2\overline x_3}\right]^{1/2}
\sum_{\mu_1\mu_2\mu_3}{D}^{1/2\,*}_{\mu_1\lambda_1}(R_{cf}({\vec{\tilde k}}_1))
{D}^{1/2\,*}_{\mu_2\lambda_2}(R_{cf}({\vec{\tilde k}}_2))
\nonumber\\
& & \quad\qquad\times {D}^{1/2\,*}_{\mu_3\lambda_3}(R_{cf}({\vec{\tilde k}}_3))
\,
\Psi_\lambda^{[c]}(\vec{k}_1,\vec{k}_2,\vec{k}_3;\mu_1,\tau_1,\mu_2,\tau_2,\mu_3,\tau_3).
\\ \nonumber
\eea

Eqs.~(\ref{eq:psicf}) and (\ref{eq:psifc}) give the correct covariant
transformation linking wave functions obtained in the (canonical) instant form,
e.g. in some CQM, and the corresponding valence-quark component of the LCWF.


\section{The valence-quark contribution}

The valence-quark contribution to GPD's is obtained by specializing
Eq.~(\ref{eq:base}) to the case $N=3$, i.e.
\bea
\label{eq:valence}
F^q_{\lambda'\lambda}(\overline x,\xi,t) & = &
\frac{1}{\sqrt{1-\xi^2}}\sum_{\lambda_i\tau_i}
\sum_{j=1}^3 \delta_{s_jq}
\int[d\overline x]_3[d\vec{k}_\perp]_3\,\delta(\overline x-\overline x_j)
\nonumber\\
& & \quad\times 
\Psi^{[f]\,*}_{\lambda'}(r',\{\lambda_i\},\{\tau_i\}) 
\Psi^{[f]}_\lambda(r,\{\lambda_i\},\{\tau_i\})\Theta(\overline x_j),
\\ \nonumber
\eea
where the front-form wave function 
$\Psi^{[f]}_\lambda(r,\{\lambda_i\},\{\tau_i\})$ is related to the
corresponding wave function 
$\Psi^{[c]}_\lambda(\{\vec{k}_i\},\{\lambda_i\},\{\tau_i\})$ in the canonical
form by Eq.~(\ref{eq:psifc}). Separating the spin-isospin from the space part of
the canonical wave function, 
\be
\label{eq:separated}
\Psi^{[c]}_\lambda (\{\vec{k}_i\},\{\lambda_i\},\{\tau_i\})
= \psi(\vec{k}_1,\vec{k}_2,\vec{k}_3)
\Phi_{\lambda\tau}(\lambda_1,\lambda_2,\lambda_3,\tau_1,\tau_2,\tau_3) ,
\ee
we have
\bea
\label{eq:transform}
& & \Psi^{[f]}_\lambda (r,\{\lambda_i\},\{\tau_i\})
=  2(2\pi)^3\left[\frac{1}{M_0}\frac{\omega_1\omega_2\omega_3}
{\overline x_1\overline x_2\overline x_3}\right]^{1/2} 
\psi(\vec{k}_1,\vec{k}_2,\vec{k}_3) 
\nonumber\\
& & \qquad \times\sum_{\mu_1\mu_2\mu_3}
{D}^{1/2\,*}_{\mu_1\lambda_1}(R_{cf}(\vec{k}_1))
{D}^{1/2\,*}_{\mu_2\lambda_2}(R_{cf}(\vec{k}_2))
{D}^{1/2\,*}_{\mu_3\lambda_3}(R_{cf}(\vec{k}_3))
\nonumber\\
& & \qquad\ {}\times
\Phi_{\lambda\tau}(\mu_1,\mu_2,\mu_3,\tau_1,\tau_2,\tau_3) ,
\\ \nonumber
\eea
where the Melosh rotations are given by
\bea
{D}^{1/2}_{\lambda\mu}(R_{cf}(\vec{\tilde k})) & = &
\bra{\lambda}R_{cf}(\overline x M_0,\vec{k}_\perp)\ket{\mu} \nonumber\\
& = & \bra{\lambda}
\frac{m+\overline xM_0-i\vec{\sigma}\cdot(\hat{\vec{z}}\times\vec{k}_\perp)}
{\sqrt{(m+\overline xM_0)^2+\vec{k}_\perp^2}}\ket{\mu}.
\\ \nonumber
\eea

\subsection{Parton distributions}

In the limit $\Delta^\mu\to 0$, where $\overline x$ goes over to the parton
momentum fraction $x$, some of the GPD's reduce to the ordinary DIS parton
distributions. In particular,
\be
\label{eq:unpol}
H^q(x,0,0) = q(x), 
\ee
where $q(x)$ is the (unpolarized) quark distribution of flavor $q$. In this
limit, the Melosh rotation matrices combine to the identity matrix and the following
simple expression is obtained for the parton distribution
\be
q(x) = \sum_{j=1}^3\delta_{\tau_j\tau_q}\int
\prod_{i=1}^3 d\vec{k}_i \, \delta\left(\sum_{i=1}^3 \vec{k}_i\right)
\,\delta\left(x-\frac{k^+_j}{M_0}\right)
\vert\Psi_\lambda^{[c]}(\{\vec{k}_i;\lambda_i,\tau_i\})\vert^2.
\ee
This expression agrees with that given, e.g., in Refs.~\cite{brodsky98,BL} and
automatically fulfills the support condition, vanishing outside the support
region $0\le x\le 1$. It also satisfies the particle number sum rule, 
\be
\int dx \,q(x) = N_q,
\ee
where $N_q$ is the number of valence quarks of flavor $q$, as well as the
momentum sum rule
\be
\int dx\, x\, [u(x) + d(x)] = 1,
\ee
where $u(x)$ and $d(x)$ are the up and down quark distributions.


\section{Results and discussion}

In this Section we present results obtained with two constituent quark models,
i.e. the  relativistic 
hypercentral quark model of Ref.~\cite{faccioli} and the
Goldstone-boson-exchange (GBE) model of Ref.~\cite{graz}. 

\subsection{Adopted models}

The hypercentral model is based on the mass operator $M=M_0+V$, where $M_0$ is
the free mass operator, 
\be
\label{eq:freemass}
M_0 = \sum_{i=1}^3 \sqrt{\vec{k}_i^2 + m_i^2},
\ee
with $\sum_i\vec{k}_i = 0$, and $m_i$ being the constituent quark masses. The
interaction $V$ is taken of the form~\cite{Santopinto}
\be
\label{eq:hyperV}
V=-\frac{\tau}{y} +\kappa_l \,y,
\ee
where $y=\sqrt{\vec{\rho}^2 + \vec{\lambda}^2}$ is the radius of the
hypersphere in six dimensions and $\vec{\rho}$ and $\vec{\lambda}$ are the
Jacobi coordinates,
\be 
\vec{\rho}=\frac{\vec{r}_1-\vec{r}_2}{\sqrt{2}},\quad
\vec{\lambda}=\frac{\vec{r}_1+\vec{r}_2-2\vec{r}_3}{\sqrt{6}}.
\ee 
The model depends on two parameters, $\tau$ and $\kappa_l$, and is able to
reproduce the basic features of the low-lying nucleon spectrum satisfactorily in
spite of its simplicity. In the hypercentral model the
resulting nucleon wave function is a product of a space and a spin-isospin part
and is SU(6) symmetric. Technical details concerning the derivation of the
relevant formulae with this model are given in the Appendix.  

In the GBE model the same semirelativistic free mass operator $M_0$ of
Eq.~(\ref{eq:freemass}) is combined with an interaction between constituent
quarks which is the sum of a (linear) color-electric confinement term and a
hyperfine (flavor-dependent) potential provided by the possible exchange of all
mesons of the pseudoscalar octet and singlet. Such mesons are considered as
Goldstone bosons appearing in the model as a consequence of the spontaneous
breaking of chiral symmetry. The model of Ref.~\cite{graz} depends on five
parameters that are fixed by looking at the baryon spectrum which is well
reproduced up to 2 GeV with the correct orderings of the positive- and
negative-parity states in the light and strange sectors. The resulting nucleon
wave functions, without any further parameter, yield a remarkably consistent
picture of the electroweak form factors~\cite{robert}.

\subsection{Results}

The spin-averaged ($H^q$) and the helicity-flip ($E^q$) GPD's calculated in the
GBE model  for the $u$ and $d$ flavours are plotted in 
Figs.~\ref{fig:fig2}--\ref{fig:fig4} as a function of
$\overline x$ at different values of $t$ and $\xi$. They are all positive apart
from $E^d$ that is always negative. This is in agreement with the findings
within the MIT bag model~\cite{bag} and also within the chiral quark-soliton
model~\cite{petrov,penttinen} where $H^u+H^d$ and $E^u-E^d$ are leading order 
in the number of colours $N_c$, while $H^u-H^d$ and $E^u+E^d$ are subleading. 
In all cases the GPD's vanish at $\overline x =\xi$ since in our approach 
they  include the contribution of valence quarks only~\cite{diehl,diehl99}. 
They also vanish beyond $\overline x=1$ satisfying the support condition 
as already mentioned. As in the MIT 
bag
model~\cite{bag} the $\xi$ dependence at fixed $t$, Figs.~\ref{fig:fig3} and
\ref{fig:fig4}, turns out to
be weak, the main effect being a small shift towards larger $\overline
x$ with increasing $\xi$.  
In addition an interesting feature emerges from Fig.~\ref{fig:fig2}: the large 
$t$-independence of both $E^q$ and $H^q$ at fixed $\xi$ in the region $x>0.5$.
Such a result is not consistent with the $t$ dependence simply factorized 
in terms of nucleon form factors~\cite{marc}.
 Actually, only the first moments of the GPD's are related to the nucleon 
elastic 
form factors~\cite{ji78}, i.e.
\be
\int_{-1}^1dx H^q(x,\xi,t) = F^q_1(t), \quad 
\int_{-1}^1dx E^q(x,\xi,t) = F^q_2(t),
\label{eq:ff_sumrule}
\ee
where $F^q_1(t)$ and $F^q_2(t)$ are the contribution of quark $q$ to the Dirac
and Pauli form factors. 
This property holds in the present approach,  for $\xi=0$ only. 
 The $\xi$ dependence which emerges from 
Figs.~\ref{fig:fig2}--\ref{fig:fig4} in calculating the first moment of 
$E^q$ and $H^q$ is due to the fact that only the valence-quark contribution is
 taken into account. The inclusion of higher order configurations in the Fock 
space would restore the validity of Eq.~(\ref{eq:ff_sumrule}) for $\xi>0$.

It is worthwhile noticing that neglecting the effect of the Melosh
rotations in the transformation from the canonical- to the front-form wave
function, Eq.~(\ref{eq:transform}), the
helicity-flip GPD's $E^q$ would be vanishing for wave functions with
S waves only. The use of the correct transformation can
also be appreciated when looking at the parton distributions for $u$ and $d$
flavours plotted in Fig.~\ref{fig:fig5}. Here the fully covariant light-front calculation
gives contrasting results with those obtained with the prescription of
Ref.~\cite{mair} including the support and flux factor corrections. To preserve
normalization the quite substantial 
shift of the peak to lower $\overline x$ values of the light-front calculation
 is compensated by a higher tail at larger $\overline x$ which reflects the 
presence of high-momentum components in the GBE wave functions.

GPD's calculated with the hypercentral and the GBE model behave quite
similarly in spite of the fact that the hypercentral model is SU(6) symmetric,
while the GBE model is not. However, it is known that the SU(6) breaking part
in the nucleon GBE wave function is small~\cite{robert,graz}. 
In Fig.~\ref{fig:fig6}  the GPD's obtained
within the hypercentral model are shown for $\xi=0$ as a function of $t$.
In order to better appreciate the difference between the GBE and hypercentral 
CQM's, 
Fig.~\ref{fig:fig7} shows the $t$ dependence of the ratio between  
$E^q$ and the anomalous magnetic moments $k^q$ predicted by the respective 
models ($k^q=\int dx \; E^q(x,t=0,\xi=0)$). The numerical values for 
$k^q$ are listed in the caption of Fig.~\ref{fig:fig7} and discussed in 
the text later on.
\newline
\noindent
In Fig.~\ref{fig:fig8} the GPD's obtained within the hypercentral model 
are shown for $t=-0.5$ (GeV)$^2$ as a function of $\xi$.
They confirm the results obtained within the GBE model.
At large $\overline x$, where it is
known that the contribution of valence quarks is substantial and higher Fock
states are less important~\cite{diehl99}, the distributions are almost
independent of $\xi$ so that at fixed $t$ their peak position for increasing
$\xi$ is shifted to higher values of $\overline x$ by the requirement that
they have to vanish here at $\overline x=\xi$. Would higher Fock states
(including gluons and antiquarks) be included, nonvanishing GPD's should occur
for $\overline x<\xi$ with $\overline x=\xi$ as a cross-over
point (see, e.g., Ref.~\cite{petrov,penttinen}). 

Finally we discuss the integral properties of $H^q$ and $E^q$ summarized in 
Eq.~(\ref{eq:ff_sumrule}).
In Fig.~\ref{fig:fig9} the proton and neutron Pauli and Dirac form factors 
are shown as function of $-t$ in both the hypercentral and GBE models.
The $F_2$ form factor at $t=0$ gives the following values for the
nucleon anomalous magnetic moments: $k^p_{GBE}=1.20$, $k^n_{GBE}=-1.06$, 
$k^p_{hyp}=0.91$, $k^n_{hyp}=-0.82$. These results are rather far from the 
experimental values and consistent with analogous light-front calculations 
when pointlike structure of the quark is assumed~\cite{pace,ratio,Simula}.
The SU(6) breaking effects which produce deviations from the value 
$k^p/k^n=-1$ come mainly from  the Melosh rotations, 
with a small additional contribution from the quark wave functions
in the case of the GBE model.
The $t$ dependence of $F_1$ and $F_2$ is rather smooth and in both models it 
does not reproduce the experimental results: a clear limitation due to the 
assumed pointlike structure of the quarks which can be solved by the 
introduction of specific quark form factors~\cite{pace,Simula}. 

\section{Concluding remarks}

A fully covariant approach has been presented linking the overlap 
representation
of generalized parton distributions~\cite{diehl,brodsky} to the non-diagonal 
one-body density matrix
in momentum space. As a result of the correct transformation of the wave
functions from the (canonical) instant-form to the (light-cone) front-form
description, the support condition is automatically fulfilled and the particle
number and momentum sum rules are also satisfied. The method has been applied 
to
the case of nucleon wave functions involving only the three valence quarks in
order to study GPD's in terms of constituent quark model wave functions. This
implies that the discussion of GPD's has to be confined to the region $\xi\le
\overline x\le 1$. In this region one can easily derive the usual parton
distributions and appreciate the advantages of a fully covariant treatment
avoiding problems connected with the support condition and the flux factor
introduced in other approaches. In addition, the helicity flip GPD's naturally
arise thanks to the relativistic effects of the Melosh rotations. However,
the limitation $\xi\le\overline x\le 1$ prevents the possibility of testing the
reduction formula of the first moment of the GPD's leading to the nucleon form
factors for $\xi\ne 0$.

Results have been presented for the hypercentral model of Ref.~\cite{faccioli}
and the Goldstone-boson-exchange model of 
Ref.~\cite{graz}. Quite similar
results are obtained within the two models, the SU(6) breaking contribution
present in the GBE wave functions being rather small. A strong $t$ and a weak
$\xi$ dependence has been found in all cases, confirming the results obtained
within the
MIT bag model~\cite{bag}. 
Keeping in mind the limitation $\xi\le\overline x\le 1$ 
due to the inclusion of the lowest order Fock-space components with 
three valence quarks only, the results obtained show that all the 
phenomenology 
for large $\overline x$ and small $t$ can be studied within the present 
approach.

\section*{Acknowledgements}%
We acknowledge M. Diehl for useful comments.


\appendix
\section{Appendix}

The structure of the nucleon wave function in the hypercentral quark
model of Ref.~\cite{faccioli} is given by Eq.~(\ref{eq:separated}) where a
product is assumed of a symmetric function of the momenta with a symmetric
function of the
spin-isospin variables as in Eq.~(\ref{eq:separated}). The spin function
$\Phi_\lambda^{S_{12}}(\lambda_1,\lambda_2,\lambda_3)$ is symmetric or
antisymmetric under the interchange of quark 1 and quark 2 for the total spin of
the pair $S_{12}=1$ and $S_{12}=0$, respectively. The third quark spin is then
coupled to $S_{12}$ to obtain the nucleon spin $\oneh$. The same is done for
isospin according to SU(2) symmetry. The resulting spin-isospin
function,
\bea
& & \Phi_{\lambda\tau}(\lambda_1,\lambda_2,\lambda_3,\tau_1,\tau_2,\tau_3)
\nonumber\\ 
& & \quad=
\frac{1}{\sqrt{2}}\left[\Phi^0_\lambda(\lambda_1,\lambda_2,\lambda_3)
\Phi^0_\tau(\tau_1,\tau_2,\tau_3) +
\Phi^1_\lambda(\lambda_1,\lambda_2,\lambda_3)\Phi^1_\tau(\tau_1,\tau_2,\tau_3)
\right],
\\ \nonumber
\eea
is then fully symmetric under all permutations. The space part of the nucleon
wave function is taken with total orbital momentum $L=0$ and is written in
momentum representation as
\be
\psi(\vec{k}_1,\vec{k}_2,\vec{k}_3) =\psi_{00}(\tilde y)Y^{(0,0)}_{[0,0,0]}(\Omega),
\ee
where $\psi_{\gamma,\nu}(\tilde y)$ is the hyperradial wave function solution in
momentum space of an
eigenvalue problem for the mass operator $M$ as a sum of the free mass operator
(\ref{eq:freemass}) and the hypercentral potential given by Eq.~(\ref{eq:hyperV}). The
hyperspherical harmonics $Y^{(L,M)}_{[\gamma,l_\rho,l_\lambda]}$ are defined on
the hypersphere of unit radius.

Such wave function is transformed to a front-form wave function according to
Eq.~(\ref{eq:transform}) and inserted into Eq.~(\ref{eq:valence}). The summation
over isospin variables gives
$\delta_{T_{12}0}\,\delta_{\tau_3 1/2} +
\delta_{T_{12}1}[\delta_{\tau_3 1/2} + 2\delta_{\tau_3-1/2}]/3$ for the
proton and 
$\delta_{T_{12}0}\,\delta_{\tau_3-1/2} +
\delta_{T_{12}1}[2\delta_{\tau_3 1/2} + \delta_{\tau_3 -1/2}]/3$ for the
neutron.

Summation over spin variables requires much more effort because of the presence
of the Melosh rotations but is straightforward along the lines of
Ref.~\cite{chung}. For the proton we can finally rewrite Eq.~(\ref{eq:valence})
as
\bea
F^q_{\lambda'\lambda} & = &\frac{3}{2}\frac{1}{\sqrt{1-\xi^2}}\frac{1}{(16\pi^3)^2}
\int\prod_{1=1}^3 d\overline x_i\,\delta\left(1-\sum_{i=1}^3\overline x_i\right)
\,\delta(\overline x - \overline x_3) \nonumber\\
& & \quad \times\int \prod_{i=1}^3
d^2\vec{k}_{\perp,i}\,\delta\left(\sum_{i=1}^3\vec{k}_{\perp,i}\right) \,
\tilde\psi^*(\{y'_i\},\{\vec{\kappa}'_{\perp,i}\})\,
\tilde\psi(\{y_i\},\{\vec{\kappa}_{\perp,i}\}) 
\nonumber\\
& & \quad \times\delta_{\tau_q\tau_3}\left\{
X^{00}_{\lambda'\lambda}(\vec{\tilde{\kappa}}',\vec{\tilde{\kappa}})
\,\delta_{\tau_31/2} + \onet
X^{11}_{\lambda'\lambda}(\vec{\tilde{\kappa}}',\vec{\tilde{\kappa}})
[\delta_{\tau_31/2} + 2\delta_{\tau_3-1/2}]\right\},
\\ \nonumber
\eea
where
\be
\tilde\psi(\{x_i\},\{\vec{\kappa}_{\perp,i}\}) =
\left[\frac{1}{M_0}
\frac{\omega_1\omega_2\omega_3}{x_1 x_2 x_3}\right]
\psi(\vec{k}_1,\vec{k}_2,\vec{k}_3),
\ee
\be
X^{00}_{\lambda'\lambda}(\vec{\tilde{k}}',\vec{\tilde{k}}) 
= \prod_{i=1}^3 N^{-1}(\vec{\tilde{k}}'_i) N^{-1}(\vec{\tilde{k}}_i)
\bra{\lambda'}A_3 +i\vec{B}_3\cdot\vec{\sigma}\ket{\lambda} (A_1A_2 +
\vec{B}_1\cdot\vec{B}_2),
\ee
\bea
X^{11}_{\lambda'\lambda}(\vec{\tilde{k}}',\vec{\tilde{k}}) 
& = &\prod_{i=1}^3 N^{-1}(\vec{\tilde{k}}'_i) N^{-1}(\vec{\tilde{k}}_i)
\onet\sum_{jj'} \Big[ (A_1A_2 -\vec{B}_1\cdot\vec{B}_2)\delta_{jj'} 
\nonumber \\
& & \ {} +
B_{1,j}B_{2,j'} + B_{1,j'}B_{2,j} + \sum_k(A_1B_{2,k} +
A_2B_{1,k})\epsilon_{j'jk}\Big]\nonumber\\
& & \quad\times
\bra{\lambda'}A_3\delta_{jj'} +\sum_k\epsilon_{j'jk}B_{3,k} 
+ i\Big(\sum_k A_3\epsilon_{j'jk}\sigma_k \nonumber\\
& & \qquad + \sigma_{j'}B_{3,j} + \sigma_jB_{3,j'}
-\vec{\sigma}\cdot\vec{B}_3\delta_{jj'}\Big)\ket{\lambda},
\\ \nonumber
\eea
with
\be
N(\vec{\tilde{k}}) = [(m+ xM_0)^2 + \vec{k}^2_\perp]^{1/2},
\ee
\be
A_i = (m+ x'_iM'_0)(m+ x_i  M_0) + k'_y k_y + k'_x k_x,
\ee
\be
B_{i,x} = - (m+ x'_iM'_0) k_y + (m+ x_i  M_0) k'_y,
\ee
\be
B_{i,y} = (m+ x'_iM'_0) k_x -  (m+ x_i  M_0) k'_x,
\ee
\be
B_{i,z} = k'_x k_y - k'_y k_x .
\ee
In the above equations the primed (unprimed) kinematical variables refer to the
final (initial) nucleon state (see Eqs.~(\ref{eq:finalconf}) and
(\ref{eq:initialconf})) and:
\be
M_0 =\sum_i\sqrt{\vec{k}^2_i+m^2_i}, \quad 
M'_0 = \sum_i\sqrt{\vec{k}'{}^2_i +m^2_i}.
\ee

In the GBE model the nucleon wave functions are expanded on a basis where the
spin-isospin part is combined with a space part in the form of correlated
gaussian functions of the Jacobi coordinates referring to a particular
partition. The total wave function is a symmetrized linear combination of such
basis functions over the three possible partitions thus ultimately violating
SU(6) symmetry. Specifically, we have
\be
\label{eq:gbewave}
\Psi^{[c]}_\lambda (\{\vec{k}_i\},\{\lambda_i\},\{\tau_i\})
= \sum_n c_n \Psi_{\alpha_n}(\{\vec{k}_i\},\{\lambda_i\},\{\tau_i\}),
\ee
where each $\Psi_{\alpha}$ is summed over all possible partitions, i.e.
\be
\Psi_{\alpha} = \Psi_{\alpha}(1,23) + \Psi_{\alpha}(2,31) +
\Psi_{\alpha}(3,12).
\ee
In turn, for each partition $(k,pq)$
\be
\Psi_{\alpha}(k,pq) = \left[\psi_{LM}(\vec{x}_k,\vec{y}_k)\bigotimes
\Phi_{\lambda}(\lambda_p,\lambda_q,\lambda_k)\right]_J
\Phi_{\tau}(\tau_p,\tau_q,\tau_k), 
\ee
where 
\be
\psi_{LM}(\vec{x}_k,\vec{y}_k) = x_k^{2\nu+\lambda}\,y_k^{2n+l}\,e^{-\beta x_k^2
-\delta y_k^2 +\gamma \vec{x}_k\cdot\vec{y}_k}\,
Y^{LM}_{\lambda l}(\hat{{x}}_k,\hat{{y}}_k),
\ee
with $2\nu+\lambda+2n+l=2N+L$, $N$ being the principal quantum number, and
\be
Y^{LM}_{\lambda l}(\hat{{x}}_k,\hat{{y}}_k) = \sum_{\mu m} (\lambda l
\mu m\vert LM) Y_{\lambda\mu}(\hat{{x}}_k) Y_{lm}(\hat{{y}}_k).
\ee
In the above equations $\vec{x}_k,\vec{y}_k$ are the Jacobi coordinates of
partition $k$. The parameters $\beta$, $\delta$ and $\gamma$ are fixed by
solving the eigenvalue problem according to, e.g., the stochastic variational
method~\cite{Suzuki:1998aa} as in Ref.~\cite{graz}. For the nucleon, $N=L=0$.

The calculation with the GBE wave function requires repeating the same steps
as with the hypercentral wave functions
for each partition of the partial contribution to
the total initial (final) nucleon wave function (\ref{eq:gbewave}).


\clearpage

\begin{figure}[ht]
\begin{center}
\epsfig{file=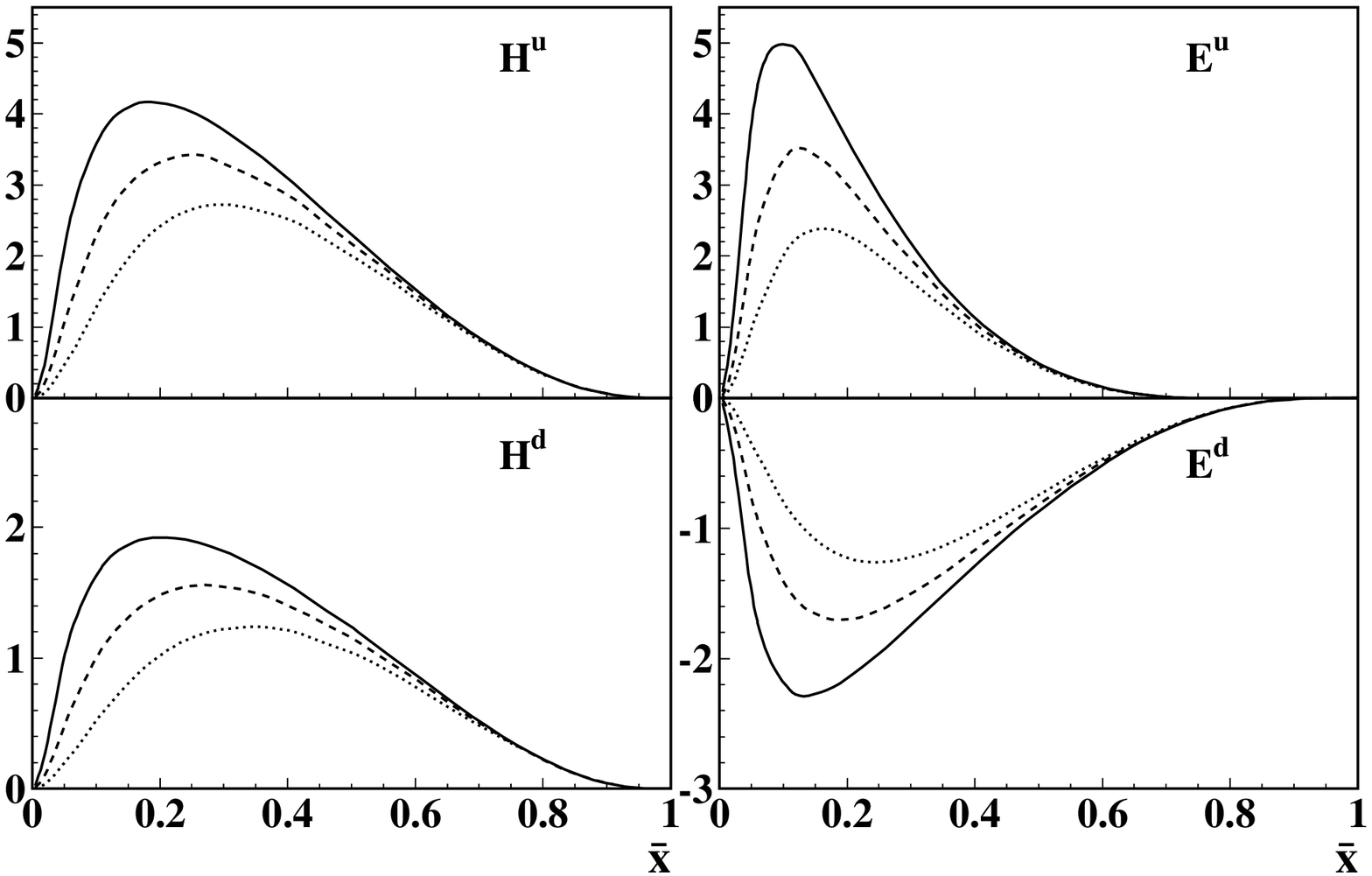,  width=33 pc, height=19 pc}
\end{center}
\caption{\small The spin-averaged ($H^q$, left panels) and the helicity-flip ($E^q$,
right panels) generalized parton distributions calculated in the GBE model for
flavours $u$ (upper panels) and $d$ (lower panels), at $\xi=0$ and different
values of $t$: $t=0$ (solid curves), $t=-0.2$ (GeV)$^2$ (dashed curves), $t=-0.5$
(GeV)$^2$ (dotted curves).}
\label{fig:fig2}
\end{figure}

\begin{figure}[ht]
\begin{center}
\epsfig{file=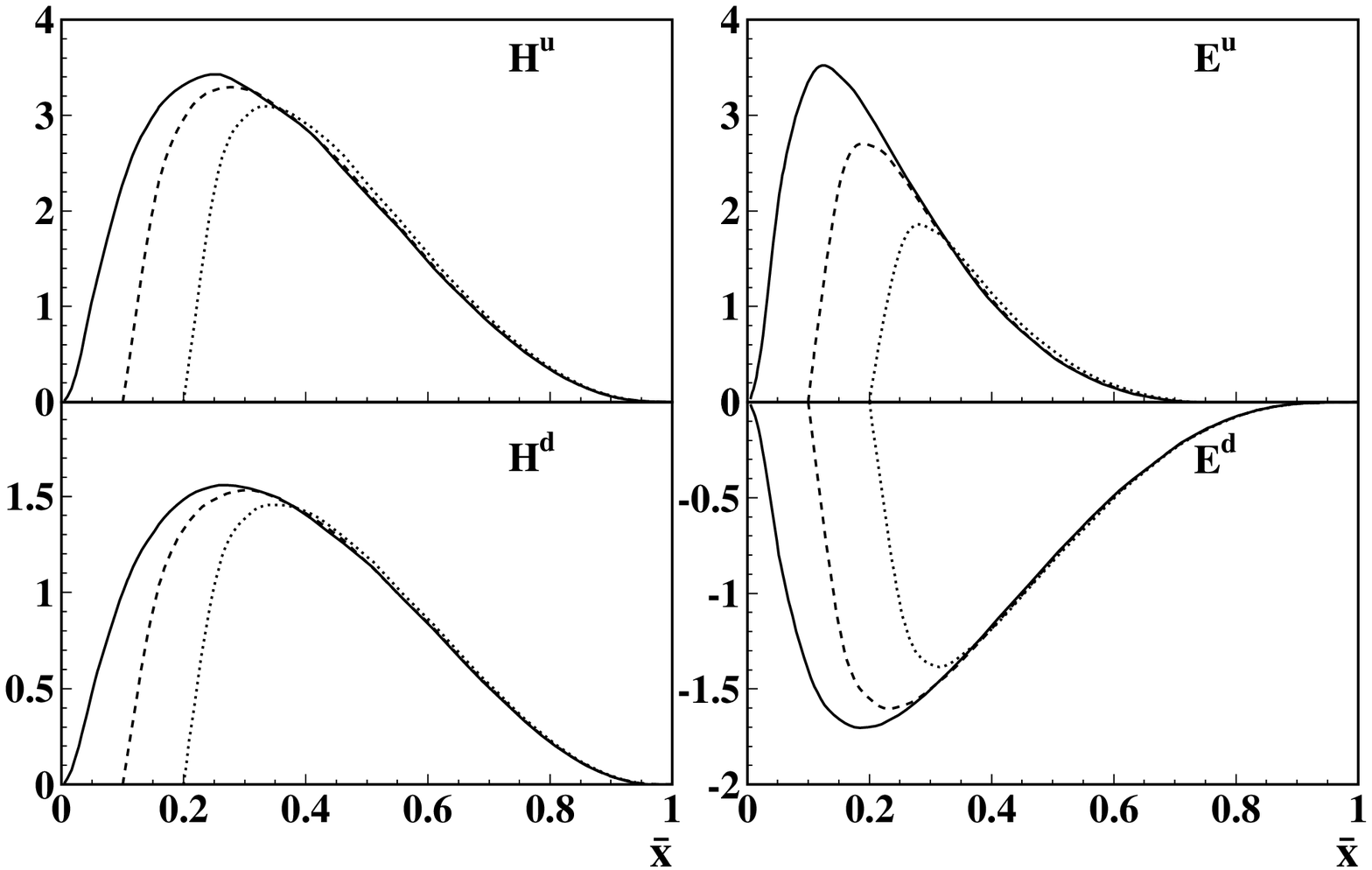,  width=33 pc, height=19 pc}
\end{center}
\caption{\small The same as in Fig. 2, but for fixed $t=-0.2$ (GeV)$^2$ and 
different values of $\xi$: $\xi =0$ (solid curves), 
$\xi=0.1$ (dashed curves), $\xi=0.2$ (dotted curves).}
\label{fig:fig3}
\end{figure}

\begin{figure}[ht]
\begin{center}
\epsfig{file=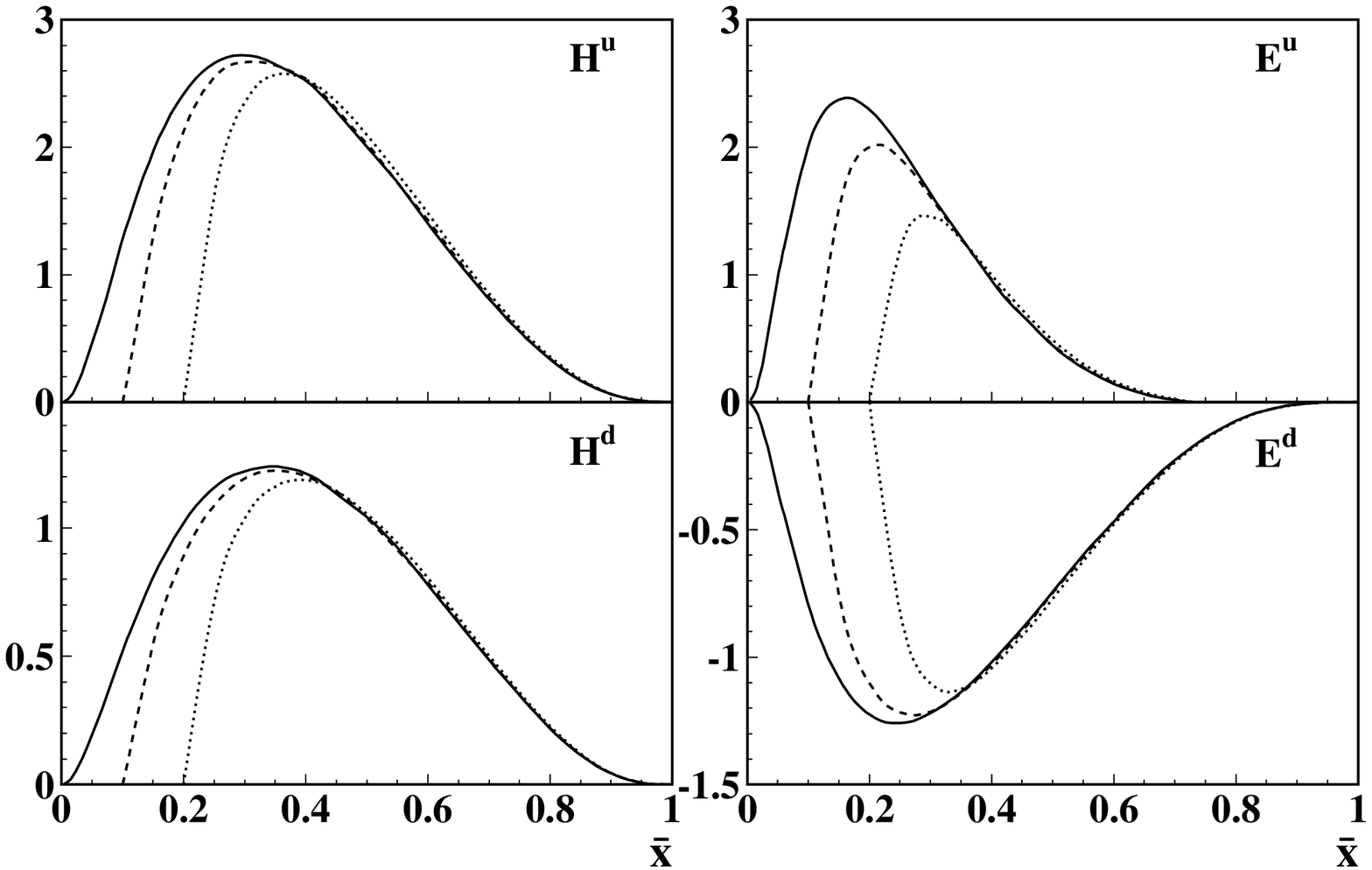,  width=33 pc, height=19 pc}
\end{center}
\caption{\small The same as in Fig. 2, but for fixed $t=-0.5$ (GeV)$^2$ and different
values of $\xi$: $\xi =0$ (solid curves), $\xi=0.1$ (dashed curves), $\xi=0.2$
(dotted curves).}
\label{fig:fig4}
\end{figure}

\begin{figure}[ht]
\begin{center}
\epsfig{file=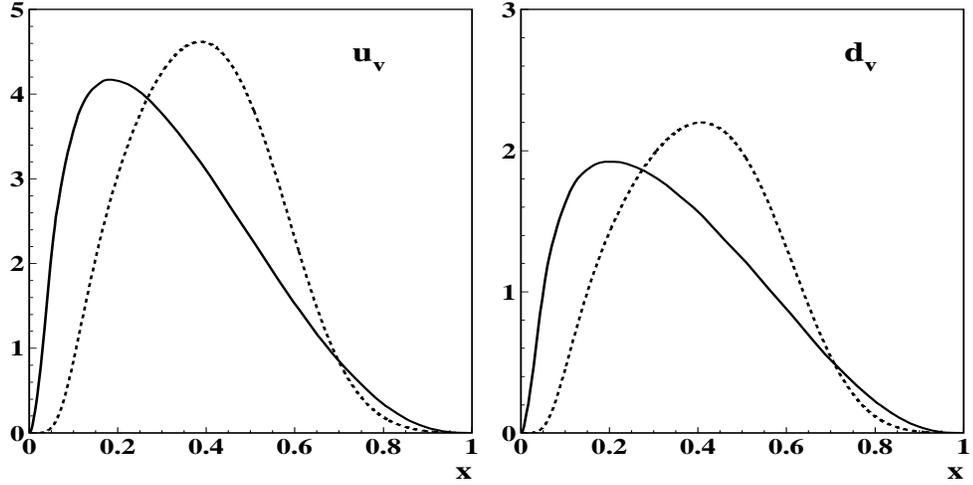,  width=31 pc, height=16 pc}
\end{center}
\caption{\small Parton distributions calculated in the GBE model for flavours $u$
(left panel) and $d$ (right panel) within the light-cone approach (solid
curves) and the approach of Ref.~\cite{mair} with the inclusion
of the support and flux factor corrections (dashed curves).}
\label{fig:fig5}
\end{figure}

\begin{figure}[ht]
\begin{center}
\epsfig{file=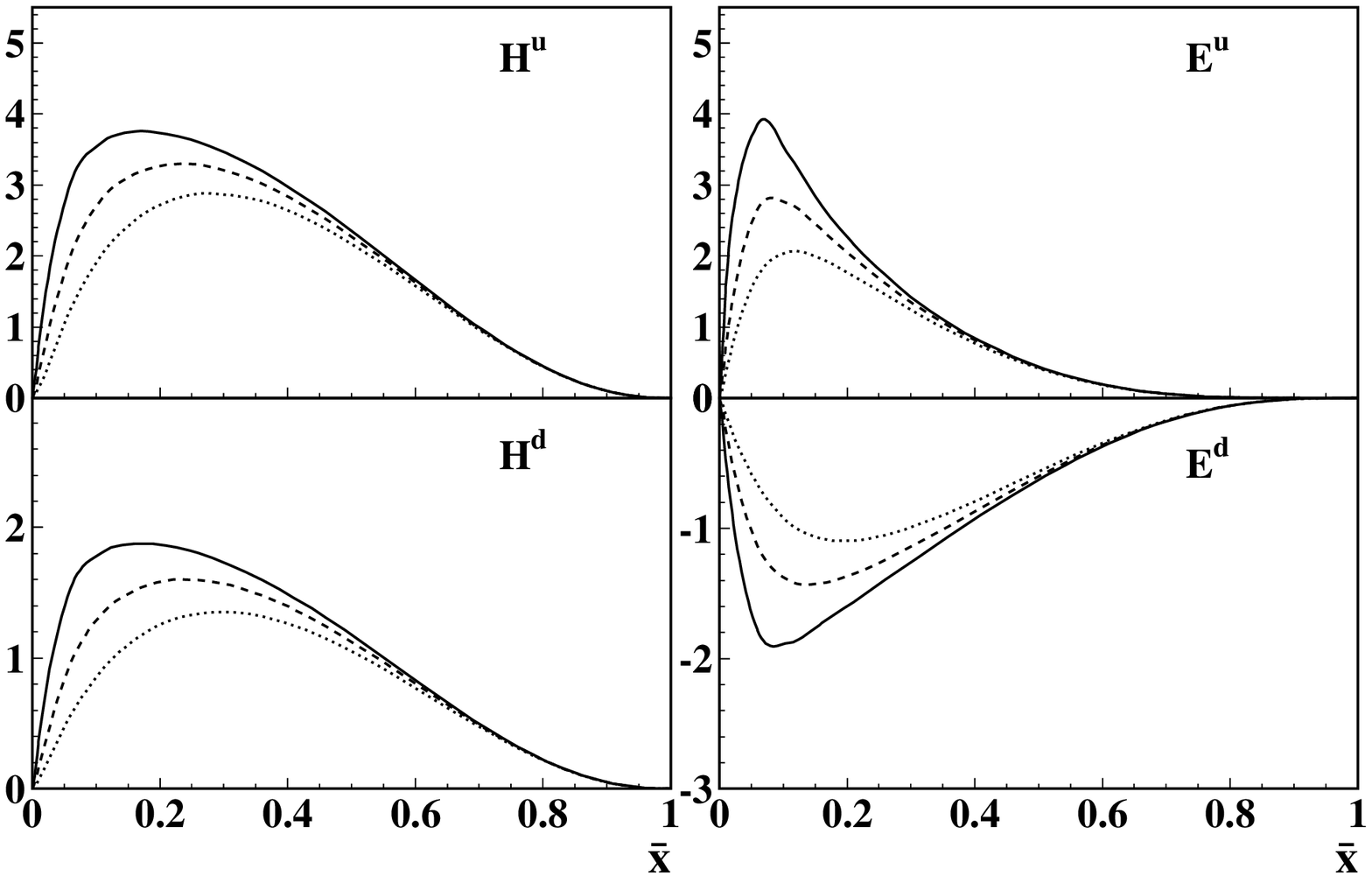,  width=33 pc, height=19 pc}
\end{center}
\caption{\small The same as in Fig.~\ref{fig:fig2}, but for the hypercentral
model.} 
\label{fig:fig6}
\end{figure}

\begin{figure}[ht]
\begin{center}
\epsfig{file=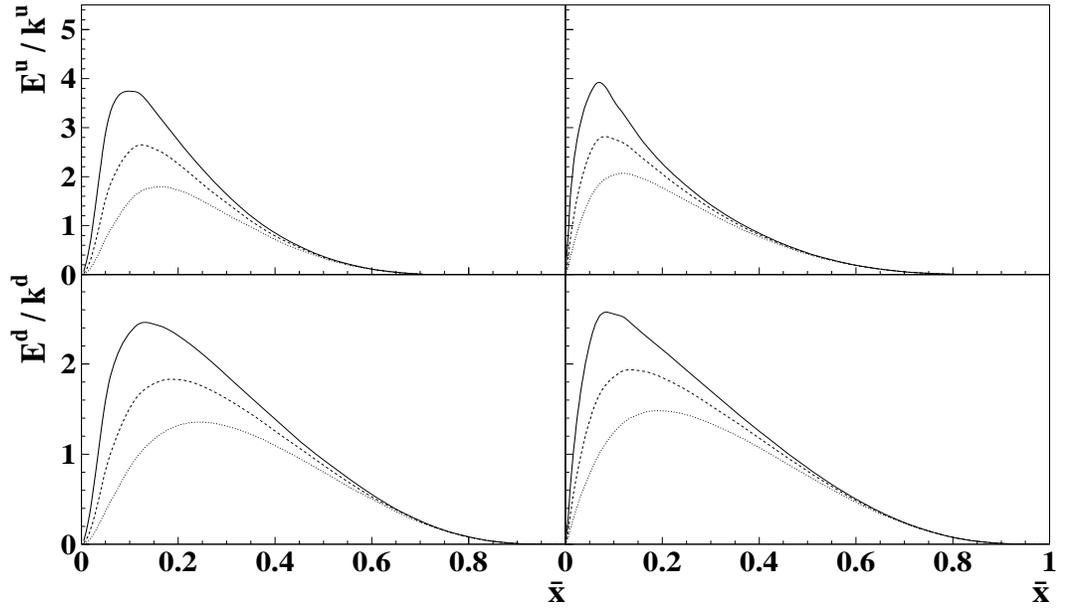,   width=33 pc, height=19 pc}
\end{center}
\caption{\small 
The ratio of the helicity-flip generalized parton distributions 
$E^q$ to the anomalous magnetic moment $k^q$ calculated 
in  the GBE model (left panels) and in the hypercentral model (right panels)
for flavours $u$ (upper panels) and $d$ (lower panels), 
at $\xi=0$ and different values of $t$: 
$t=0$ (solid curves), $t=-0.2$ (GeV)$^2$ (dashed curves), $t=-0.5$ (GeV)$^2$ 
(solid curves).
The values of the quark anomalous magnetic moments predicted from the two 
models are:
$k^u_{GBE}=1.33$, $k^d_{GBE}=-0.93$, $k^u_{hyp}=1.00$, $k^u_{hyp}=-0.74$. }
\label{fig:fig7}
\end{figure}

\begin{figure}[ht]
\begin{center}
\epsfig{file=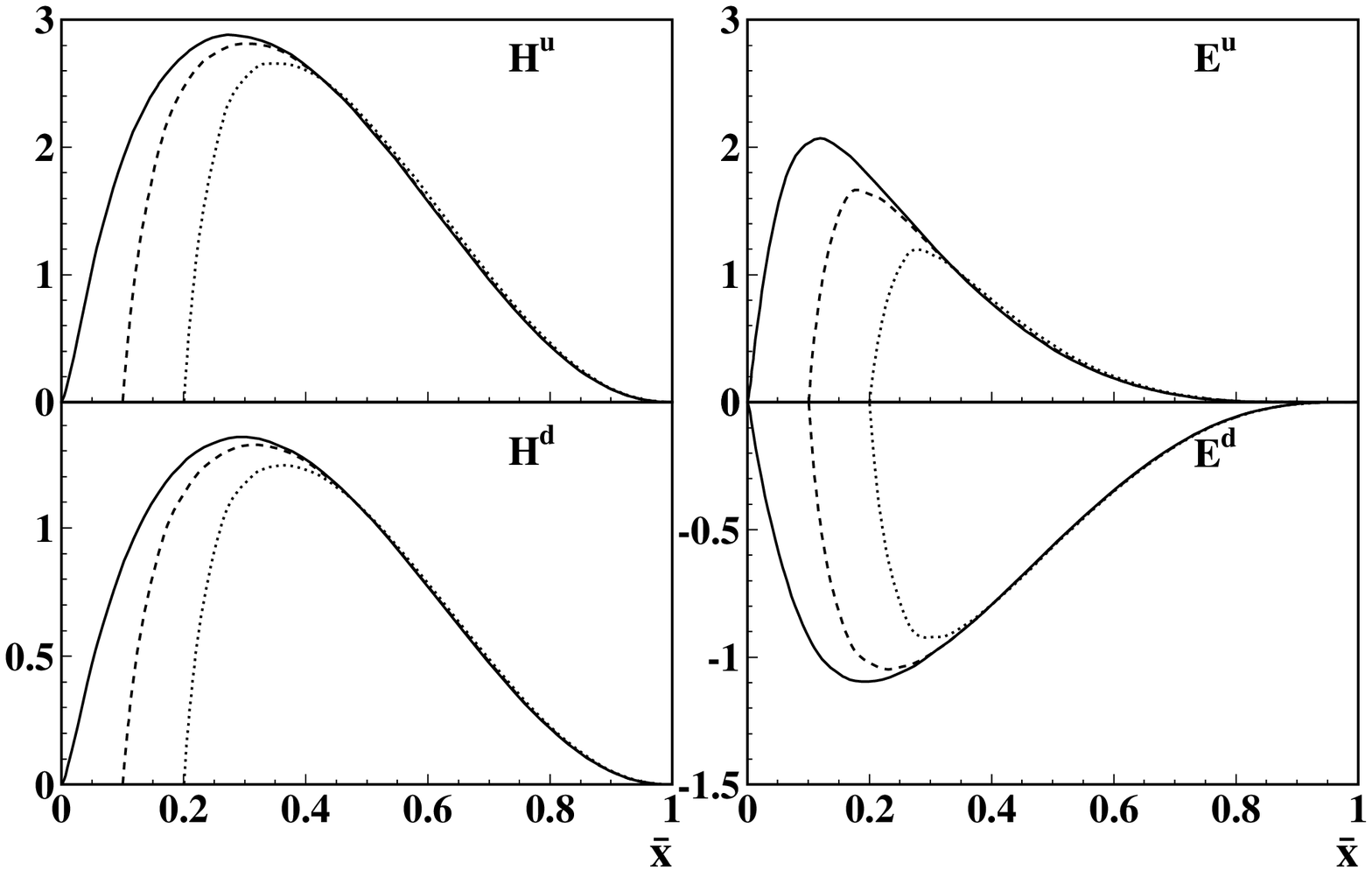,  width=33 pc, height=19 pc}
\end{center}
\caption{\small The same as in Fig.~\ref{fig:fig4}, but for the hypercentral
model.} 
\label{fig:fig8}
\end{figure}

\begin{figure}[ht]
\begin{center}
\epsfig{file=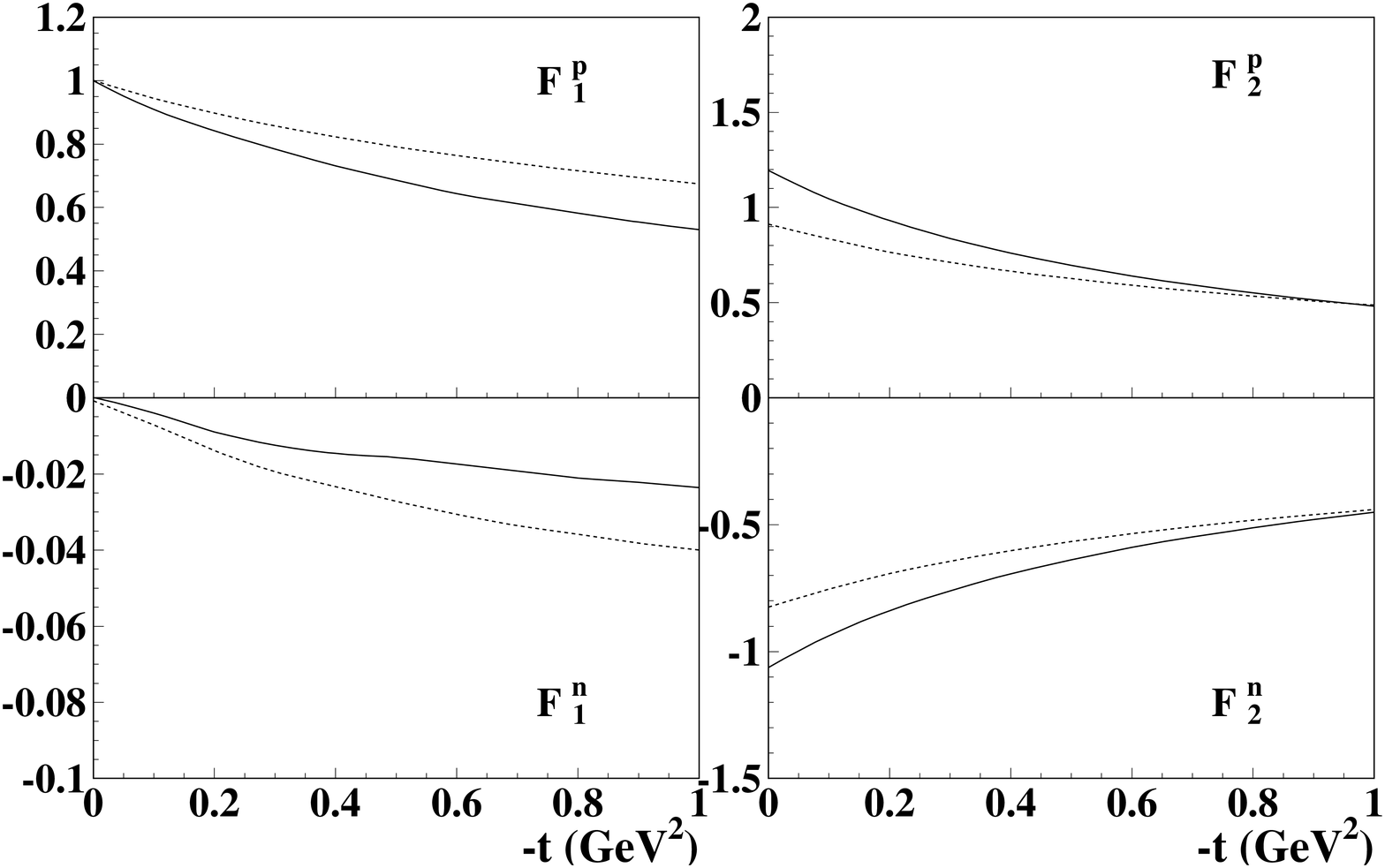,  width=33 pc, height=19 pc}
\end{center}
\caption{\small 
The Pauli (left panels) and Dirac (right panels) form factors
calculated in the hypercentral model (dashed curves) and in the GBE model 
(solid curves) 
for the proton (upper panels) and the neutron (lower panels).}
\label{fig:fig9}
\end{figure}

\clearpage


\begin{thebibliography} {99}

\bibitem{muller}
D. M\"uller, D. Robaschik, B. Geyer, F.M. Dittes, J. Ho\u rej\u si, Fortsch.
Phys. 42 (1994) 101.

\bibitem{radyushkin96a}
A.V. Radyushkin, Phys. Lett. B 380 (1996) 417.

\bibitem{radyushkin96b}
A.V. Radyushkin, Phys. Lett. B 385 (1996) 333.

\bibitem{ji78}
Xiangdong Ji, Phys. Rev. Lett. 78 (1997) 610.

\bibitem{radyushkin97}
A.V. Radyushkin, Phys. Rev. D 56 (1997) 5524.

\bibitem{diehl01}
M. Diehl, Eur. Phys. J. C 19 (2001) 485.

\bibitem{jiosborne}
X. Ji and J. Osborne, Phys. Rev. D 58 (1998) 094018.

\bibitem{collins99}
J.C. Collins and A. Freund, Phys. Rev. D 59 (1999) 074009.

\bibitem{ji55}
Xiangdong Ji, Phys. Rev. D 55 (1997) 7114.

\bibitem{belitsky}
A.V. Belitsky, D. M\"uller and A. Kirchner, Nucl. Phys. B 629 (2002) 323.

\bibitem{collins97}
J.C. Collins, L.L. Frankfurt and M. Strikman, D 56 (1997) 2982.

\bibitem{jig}
Xiangdong Ji, J. Phys. G 24 (1998) 1181.

\bibitem{radyushkin01}
A.V. Radyushkin, hep-ph/0101225.

\bibitem{goeke}
K. Goeke, M.V. Polyakov and M. Vanderhaeghen, Progr. Part. Nucl. Phys. 47
(2001) 401.

\bibitem{guidal}
M. Guidal, Nucl. Phys. A699 (2002) 200.

\bibitem{nowak}
V.A. Korotkov and W.-D. Nowak, Eur. Phys. J. C 23 (2002) 455.

\bibitem{radyushkin99}
A.V. Radyushkin, Phys. Rev. D 59 (1999) 014030.

\bibitem{marc}
M.Vanderhaeghen, P.A.M. Guichon and M. Guidal, Phys. Rev. D 60 (1999) 094017.

\bibitem{vanderhaeghen}
M. Vanderhaeghen, Eur. Phys. J. A 8 (2000) 455.

\bibitem{bag}
X. Ji, W. Melnitchouk and X. Song, Phys. Rev. D 56 (1997) 5511.

\bibitem{petrov}
V.Yu. Petrov, P.V. Pobylitsa, M.V. Polyakov, I. B\"ornig, K. Goeke and C. Weiss,
Phys. Rev. D 57 (1998) 4325.

\bibitem{penttinen}
M. Penttinen, M.V. Polyakov and K. Goeke, Phys. Rev. D 62 (2000) 014024.

\bibitem{diehl}
M. Diehl, Th. Feldmann, R. Jakob and P. Kroll, Nucl. Phys. B 596 (2001) 33.

\bibitem{brodsky}
S.J. Brodsky, M. Diehl and D.S. Hwang, Nucl. Phys. B 596 (2001) 99.

\bibitem{diehl99}
M. Diehl Th. Feldmann, P. Kroll and R. Jakob, Eur. Phys. J. C 8 (1999) 409.

\bibitem{scopetta}
S. Scopetta and V. Vento, hep-ph/0201265.

\bibitem{faccioli}
P. Faccioli, M. Traini and V. Vento, Nucl. Phys. A 656 (1999) 400.


\bibitem{conci}
L. Conci and M. Traini, Few-Body Systems, 8, 123 (1990); M. Traini, L. Conci and
U. Moschella, Nucl. Phys. A 544 (1992) 731.

\bibitem{thomas}
F.M. Steffens and A.W. Thomas, Nucl. Phys. A 568 (1994) 798.

\bibitem{mair}
M. Traini, V. Vento, A. Mair and A. Zambarda, Nucl. Phys. A 614 (1997) 472; A.
Mair and M. Traini, Nucl. Phys. A 624 (1997) 564.

\bibitem{keister}
B.D. Keister and W.N. Polyzou, Adv. Nucl. Phys. 20 (1991) 225.

\bibitem{BT}
B. Bakamjian and L.H. Thomas, Phys. Rev. 92 (1953) 1300.

\bibitem{pace}
 F. Cardarelli, E. Pace, G. Salm\`e, and S. Simula, Phys. Lett B 357 (1995) 
267.
\bibitem{robert}
R.F. Wagenbrunn, S. Boffi, W. Klink, W. Plessas and M. Radici, Phys. Lett. B
511 (2001) 33; L.Ya. Glozman, M. Radici, R.F. Wagenbrunn, S. Boffi and W.
Plessas, Phys. Lett. B 516 (2001) 183; S. Boffi, L.Ya. Glozman, W. Klink, W.
Plessas, M. Radici and R.F. Wagenbrunn, Eur. Phys. J. A 14 (2002) 17-21.

\bibitem{graz}
L.Ya. Glozman, W. Plessas, K. Varga and R.F. Wagenbrunn, Phys. Rev. D 58
(1998) 094030.

\bibitem{dziemb}
Z. Dziembowski, Phys. Rev. D 37 (1988) 768.

\bibitem{brodsky98}
S.J. Brodsky, H.-Ch. Pauli and S.S. Pinsky, Phys. Rep. 301 (1998) 299.

\bibitem{BL}
S.J. Brodsky and G.P. Lepage, in: A.H. Mueller (ed.), {\sl Perturbative Quantum
Chromodynamics}, World Scientific, Singapore, 1989, p.93. 

\bibitem{Santopinto}
The hypercentral-potential model has been introduced, within a nonrelativistic
framework, in:
M. Ferraris, M.M. Giannini, M. Pizzo, E. Santopinto, L. Tiator, 
Phys. Lett. B 364 (1995) 231.

\bibitem{ratio}
B. Pasquini, M. Traini and S. Boffi, Phys. Rev. D 65 (2002) 074028.


\bibitem{Simula}
S. Simula, in: D. Drechsel and L. Tiator (eds.), {\sl NSTAR 2001, Proceedings
of the Workshop on the Physics of Excited Nucleons}, World Scientific, 
Singapore, 2001, p.135, and  nucl-th/0105024

\bibitem{chung}
P.L. Chung and F. Coester, Phys. Rev. D 44 (1991) 229.

\bibitem{Suzuki:1998aa}
Y.~Suzuki and K.~Varga, \emph{Stochastic Variational Approach to
Quantum-Mechanical Few-Body Problems}, (Springer Verlag, Berlin/Heidelberg,
1998). 

\end{thebibliography}
\end{document}